\documentclass[twoside]{article}
\usepackage{amsfonts}
\usepackage{amssymb}
\usepackage{qic}

\begin{document}
\setlength{\textheight}{8.0truein}    

\runninghead{How a Clebsch-Gordan Transform Helps to Solve  $\ldots$}
            {D. Bacon}

\normalsize\textlineskip
\thispagestyle{empty}
\setcounter{page}{1}


\vspace*{0.88truein}

\alphfootnote

\fpage{1}

\centerline{\bf
How a Clebsch-Gordan Transform Helps to Solve}
\centerline{\bf
the Heisenberg Hidden Subgroup
Problem}
\vspace*{0.035truein}
\centerline{\footnotesize
D. Bacon}
\vspace*{0.015truein}
\centerline{\footnotesize\it Department of Computer Science \& Engineering,
University of Washington, Box 352350}
\baselineskip=10pt
\centerline{\footnotesize\it Seattle, WA 98109,
USA}
\vspace*{10pt}

\vspace*{0.21truein}

\abstracts{
It has recently been shown that quantum computers can efficiently solve the
Heisenberg hidden subgroup problem, a problem whose classical query complexity
is exponential.  This quantum algorithm was discovered within the framework of
using pretty good measurements for obtaining optimal measurements in the hidden
subgroup problem.  Here we show how to solve the Heisenberg hidden subgroup
problem using arguments based instead on the symmetry of certain hidden
subgroup states.  The symmetry we consider leads naturally to a unitary
transform known as the Clebsch-Gordan transform over the Heisenberg group.
This gives a new representation theoretic explanation for the pretty good
measurement derived algorithm for efficiently solving the Heisenberg hidden
subgroup problem and provides evidence that Clebsch-Gordan transforms over
finite groups are a new primitive in quantum algorithm design.
}{}{}

\vspace*{10pt}

\vspace*{3pt}
\communicate{to be filled by the Editorial}

\vspace*{1pt}\textlineskip    

\section{Introduction}

In 1994 Peter Shor discovered that a quantum computer could efficiently factor
integers\cite{Shor:94a}, a problem which is widely suspected to be intractable
on a classical computer\cite{Crandall:01a}.  Since this discovery an intrepid
group of researchers have been attempting to discover quantum algorithms which
lie beyond Shor's factoring algorithm with mixed success.  On the one hand, a
great deal of success has been had in achieving polynomial speedups over
classical algorithms within the framework of Grover's quantum search
algorithm\cite{Grover:96a}. On the other hand, quantum algorithms which, like
Shor's algorithm, perform exponentially faster than the best classical
algorithms have been harder to come by.  To be sure, notable successes have
been achieved, including Hallgren's efficient quantum algorithm for solving
Pell's equation\cite{Hallgren:02a} and exponential speedups in certain quantum
random walks\cite{Childs:02b}, but so far there have been no new efficient
quantum algorithms which move quantum computing's killer application beyond
factoring integers and the resulting breaking of multiple public key
cryptosystems.

Among the most tempting problems which might be exponentially sped up on a
quantum computer is the graph isomorphism problem.  The reasons for this are
two-fold.  First, graph isomorphism belongs to a complexity class very much
like that which contains integer factoring.  In particular, the decision
version of factoring is known to be in the complexity $NP \cap coNP$ while
graph isomorphism is known to be in the similar complexity class $NP \cap
coAM$.  Algorithms in these complexity classes are unlikely to be
$NP$-complete.  Further, both integer factoring and graph isomorphism are not
known to have classical polynomial time algorithms despite considerable effort
to find such algorithms.  Thus graph isomorphism is, like integer factoring,
of Goldilocks-like classical complexity, not too hard such that efficiently
solving it would revolutionize our notion of tractable, but not so easy as to
have already fallen into $P$ (or $BPP$.)  The second reason for attempting to
find efficient quantum algorithms for the graph isomorphism problem is that
this problem can be solved if there was an
efficient quantum algorithm for the non-Abelian version of the problem which
lies at the heart of Shor's algorithm, the hidden subgroup
problem\cite{Boneh:95a,Ettinger:99b}.  Thus motivated, a great deal of effort
has been expended in the last few years attempting to solve the non-Abelian
hidden subgroup problem efficiently on a quantum computer.  Towards this end, a
series of efficient quantum algorithms for certain non-Abelian hidden subgroups
have been
developed\cite{Grigni:00a,Hallgren:00a,Ivanyos:01a,Friedl:03a,Moore:04a,Gavinsky:04a,Bacon:05a,Childs:05a}.
At the same time a series of negative results towards the standard approach to
solving the hidden subgroup problem on a quantum computer have also
appeared\cite{Grigni:00a,Hallgren:00a,Regev:02a,Moore:05a,Moore:05b,Bacon:06c,Hallgren:06a}.
Viewed pessimistically, these results cast doubt on whether quantum computers
can be used to efficiently solve non-Abelian hidden subgroup problems.  An
alternative optimistic view is also possible.  In this view what these results
show is that any efficient quantum algorithm for the non-Abelian hidden
subgroup problem (in what is known as the standard method) must have a particular form.  Specifically such algorithms must perform quantum circuits across many separate quantum queries to a hidden subgroup oracle.  If we are to view these results in a positive manner, then this tells us that what is needed, if we are going to efficiently solve non-Abelian hidden subgroup problems, are new quantum transforms which can act across many such queries.

In this paper we provide some evidence in favor of this optimistic view.  Recently, Bacon, Childs, and van Dam\cite{Bacon:05a} have shown that quantum computers
can efficiently solve the hidden subgroup problem for certain semidirect
product groups.  One such group which admits an efficient quantum algorithm is
the Heisenberg group, ${\mathbb Z}_p^2 \rtimes {\mathbb Z}_p$.  The algorithm
of Bacon, Childs, and van
Dam\cite{Bacon:05a} was discovered using the framework of pretty good
measurements\cite{Ip:03a,Bacon:05a,Moore:05c,Childs:05a,Bacon:06c}.  This
yields a particular algorithm for solving the Heisenberg hidden subgroup
problem which is optimal and which is related to the solution of certain
algebraic equations.  Here we show that the structure of this quantum algorithm
can be derived almost solely from symmetry arguments.  These symmetry arguments
give rise to a transform, the Clebsch-Gordan transform over the Heisenberg
group, which can be used to help efficiently solve the Heisenberg hidden
subgroup problem.  Previously a Clebsch-Gordan transform was used by Kuperberg
to find a subexponential algorithm for the dihedral hidden subgroup
problem\cite{Kuperberg:03a,Regev:04a}.  Clebsch-Gordan transforms over the
unitary group\cite{Bacon:06d,Bacon:06e} and a certain form of measurement were
demonstrated to not help in solving a hidden subgroup problem by Childs,
Harrow, and Wocjan\cite{Childs:06a}.  Further Moore, Russell, and Sniady have recently shown that a certain form of Clebsch-Gordan transform used to perform a quantum algorithm which mimics Kuperberg's algorithm cannot be used to obtain even a subexponential time algorithm for the hidden subgroup problem relevant to graph isomorphism\cite{Moore:06a}.  Here we show that these negative results can be overcome, at least for the Heisenberg group, by performing a
Clebsch-Gordan transform over the relevant finite group instead of over the
unitary group, and further, and in direct contrast to the work of Moore, Russell, and Sniady\cite{Moore:06a}, by working with a particular register, known as the multiplicity space register, which is output from a Clebsch-Gordan transform.  This is the first time a Clebsch-Gordan transform and its multiplicity register has been identified as a key component in producing a polynomial time algorithm for a hidden subgroup problem.

Our motivation for using a Clebsch-Gordan transform in the hidden subgroup
problem arises from considering a slight variant of the standard hidden
subgroup problem.  The setup for this variant is identical to the hidden
subgroup problem, but now the task is not to identify the hidden subgroup but
to return which set of conjugate subgroups the hidden subgroup belongs.  It is
this latter problem, which we call the hidden subgroup conjugacy problem, which
endows our system with extra symmetries which allow us to exploit a
Clebsch-Gordan transform.  An essential step in our use of the Clebsch-Gordan
transform is a demonstration that the hidden subgroup problem and the hidden
subgroup conjugacy problem are quantum polynomial time equivalent for the
Heisenberg group.

The outline of our paper is as follows.  In Section~\ref{sec:hsp} we introduce
the hidden subgroup problem and discuss relevant prior work on quantum
algorithms for this problem.  In Section~\ref{sec:hscp} we introduce a variant
of the hidden subgroup problem which we call the hidden subgroup conjugacy
problem.  In Section~\ref{sec:symhsp} we review arguments for why the symmetry
of hidden subgroup states leads one (in the standard approach to the hidden
subgroup problem) to perform a quantum Fourier transform over the relevant
group.  In Section~\ref{sec:symhscp} we present our first new results in
showing that for the hidden subgroup conjugacy problem, symmetry arguments
lead
one to perform (in addition to the quantum Fourier transform) a transform known as the Clebsch-Gordan transform over the relevant
group.  In Section~\ref{sec:heis} we introduce the Heisenberg group and show
how solving the hidden subgroup conjugacy problem for this group leads to an
algorithm for the hidden subgroup problem for this group.  In
Section~\ref{sec:cgheis} we discuss the Clebsch-Gordan transform over the
Heisenberg group and show how to efficiently implement this transform with a
quantum circuit.  Finally in Section~\ref{sec:algo} we put this Clebsch-Gordan
transform to use on the Heisenberg hidden subgroup conjugacy problem and show
how this allows one to efficiently solve the Heisenberg hidden subgroup
problem.

\section{The Hidden Subgroup Problem} \label{sec:hsp}

Here we define the hidden subgroup problem and give a brief history of attempts
to solve this problem efficiently on a quantum computer.

The hidden subgroup problem (HSP) is as follows.  Suppose we are given a known
group ${\mathcal G}$ and a function $f$ from this group to a set $S$,
$f:{\mathcal G} \rightarrow S$.  This function is promised to be constant and
distinct on left cosets of a subgroup ${\mathcal H}$, i.e. $f(g_1)=f(g_2)$ iff
$g_1$ and $g_2$ are members of the same left coset $g {\mathcal H}$.  We do
not
know the subgroup ${\mathcal H}$.  The goal of the HSP is
to identify the hidden subgroup ${\mathcal H}$ by querying the function $f$.
An
algorithm for the HSP is said to be efficient if the subgroup can be
identified
with an algorithm which runs polynomially in the logarithm of the size of the
group,  $O({\rm polylog}|{\mathcal G}|)$.  We will assume, throughout this
paper that the group ${\mathcal G}$ is finite and its elements, along with
elements of ${\mathcal S}$, have an efficient representation ($O({\rm
polylog}|{\mathcal G}|)$) in terms of bitstrings.

In the quantum version of the HSP we are given access to an oracle which
queries the function, $U_f$.  This oracle is assumed to have been constructed
using classical reversible gates in such a way that applying it to $|g\rangle
\otimes |0\rangle$ results in $|g\rangle \otimes |f(g)\rangle$.  In the
standard query model of the HSP, one inputs a
superposition over all group elements into the first register of the quantum
oracle and $|0\rangle$ into the second register of oracle.  This produces the
state
\begin{equation}
U_f {1 \over \sqrt{|{\mathcal G}|}} \sum_{g \in {\mathcal G}}
|g\rangle \otimes |0\rangle = {1 \over \sqrt{|{\mathcal G}|}} \sum_{g
\in {\mathcal G}} |g\rangle \otimes |f(g)\rangle.
\end{equation}
Suppose we now disregard (measure, throw away) the second register.  Due to
the
promise on $f$, the state of the first register is then a mixed state whose
exact form depends on the hidden subgroup ${\mathcal H}$,
\begin{equation}
\rho_{\mathcal H}={|{\mathcal H}| \over |{\mathcal G}|} \sum_{g={\rm
coset~representative}} |g{\mathcal H} \rangle \langle g{\mathcal H}|,
\label{eq:hsstate}
\end{equation}
where we have defined the left coset states
\begin{equation}
|g{\mathcal H}\rangle = {1 \over \sqrt{|{\mathcal H}|}} \sum_{h \in {\mathcal
H}} |gh\rangle.
\end{equation}
We will call $\rho_{\mathcal H}$ the hidden subgroup state.  In this paper we
will restrict ourselves to algorithms which use the above standard procedure
(with one slight variation of not querying over the entire group but only
querying over a subgroup of the group.)

The HSP, when the group is Abelian, can be solved
efficiently on a quantum computer\cite{Shor:94a,Kitaev:95a,Boneh:95a}.  The
vast majority of early efficient quantum algorithms which demonstrated speedups
over classical algorithms, including Simon's algorithm\cite{Simon:94a}, the
Deutsch-Jozsa algorithm\cite{Deutsch:92a}, the non-recursive
Bernstein-Vazirani algorithm\cite{Bernstein:97a}, and Shor's
algorithm\cite{Shor:94a} can all be recast as Abelian
HSPs.  Given the central nature of this problem to these
algorithms, a natural generalization was to consider the hidden subgroup
problem for non-Abelian groups.  It was quickly noted that if one could
efficiently solve the HSP for the symmetric group (or for
the wreath product group, ${\mathcal S}_n \wr {\mathcal S}_2$) then one would
immediately have an efficient algorithm for the graph isomorphism
problem\cite{Boneh:95a,Beals:97a,Ettinger:99b,Hoyer:97a}.  There is no known
efficient algorithm for the graph isomorphism problem despite a considerable
amount of effort to solve this problem classically\cite{Kobler:93a}.  Adding
considerably to the interest in the non-Abelian HSP was the discovery by
Regev\cite{Regev:02a} that a quantum polynomial time algorithm for the dihedral
hidden subgroup problem could be used to solve certain unique shortest vector
in a lattice problems.  Solving either of these two problems would represent a
significant breakthrough in quantum algorithms and thus a great deal of
research has been aimed at understanding the non-Abelian HSP.

Work on the non-Abelian HSP can be roughly divided into two categories:
progress in finding efficient new quantum algorithms for the problem and
attempts to elucidate the reason that the standard approach fails to efficiently solve the HSP.  For the former, a small, but significant amount of success has been achieved with a general trend of finding algorithms which are, loosely, more and more non-Abelian.  The latter has recently culminated in showing that for HSP relevant to the graph isomorphism problem will require a new class of measurements of measurement across multiple hidden subgroup states if an efficient quantum algorithm is possible.  Here we review the progress in both of these categories.

An early positive result for the non-Abelian hidden subgroup problem was the
discovery that the problem had an efficient quantum algorithm when the hidden
subgroups are normal and there exists an efficient quantum algorithm for the
quantum Fourier transform over the relevant group\cite{Hallgren:00a}.  Further
it was shown that there is an efficient quantum algorithm for the HSP when the
groups are ``almost Abelian'' \cite{Grigni:00a} or, a bit more generally, when
the group is ``near Hamiltonian''\cite{Gavinsky:04a} (these conditions mean
roughly that the intersection of the normalizers of all the subgroups of the
group is large.)  Groups with small commutator subgroup\cite{Ivanyos:01a} along
with solvable groups of bounded exponent and of bounded derived
series\cite{Friedl:03a} also admit HSPs which can be efficiently solved on a
quantum computer. Further, a series of efficient quantum algorithms for the
non-Abelian HSP over semidirect product groups have been discovered.  Among
these are certain groups of the form ${\mathbb Z}_{p^k}^n \rtimes {\mathbb
Z}_2$ for a fixed power of a prime $p^k$\cite{Friedl:03a}, $q$-hedral groups
with sufficiently large $q$\cite{Moore:04a}, and certain metacyclic groups as
well as groups of the form ${\mathbb Z}_p^r \rtimes {\mathbb Z}_p$ for fixed
$r$\cite{Bacon:05a}.   This last work includes an efficient quantum algorithm
for the Heisenberg HSP ($r=2$) which is the main subject of this paper.
Further, some non-Abelian groups can be solved using a classical reduction and
only the Abelian version of the HSP\cite{Bacon:05a}, including the groups
${\mathbb Z}_{2}^n \wr {\mathbb Z}_2$\cite{Rotteler:98a} and particular
semidirect products of the form ${\mathbb Z}_{p^k} \rtimes {\mathbb Z}_p$ with
$p$ an odd prime\cite{Inui:04a}.  Subexponential, but not polynomial, time
quantum algorithms for the dihedral group were discovered by
Kuperberg\cite{Kuperberg:03a} and subsequently improved to use only polynomial
space by Regev\cite{Regev:04a}.  Finally, a subexponential time quantum
algorithm for hidden subgroup problems over direct product groups was recently
discovered by Alagic, Moore, and Russell\cite{Alagic:06a}

In addition to the explicit efficient quantum algorithms for the non-Abelian
HSP given above, a great deal of work has also been performed examining the
query complexity of the problem and the related questions of what is needed in
order to information theoretically reconstruct the hidden subgroup.  A positive
result along these lines was the result of Ettinger, Hoyer, and Knill who
showed that the query complexity of the HSP is polynomial\cite{Ettinger:99a}.
Thus it is known that if one is given a polynomial number of copies of the HSP
state $\rho_{\mathcal H}$ then there exists a quantum measurement which can
correctly identify ${\mathcal H}$ with high probability.  However, no efficient
quantum algorithm implementing this measurement is known to exist, except for
the cases of the efficient quantum algorithms for the HSP described above.

Given that the query complexity of the HSP is polynomial, it is natural to ask
how tight this query complexity is.  In particular it is natural to ask how
many copies of the hidden subgroup state must be supplied in order for there to
exist a measurement on these copies which we can efficiently perform and which provides enough information to reconstruct the hidden subgroup.  In such cases we say that the hidden subgroup
can be information theoretically reconstructed.  Note, however that being
information theoretically reconstructible does not mean that there is an
efficient algorithm for the problem because the classical post processing
required to reconstruct the hidden subgroup may not be tractable.  What is
known about the number of copies needed to information theoretically construct
the hidden subgroup?  On the one hand it is known that for certain groups, in
particular for the dihedral\cite{Ettinger:00a}, affine\cite{Moore:04a}, and
Heisenberg groups\cite{Radhakrishnan:05a}, measurements on a single register of the hidden subgroup state is sufficient for information
theoretic reconstruction of the hidden subgroup, and these quantum measurements can be efficiently enacted. However, it was shown by
Moore, Russell and Schulman\cite{Moore:05a} that for the particular case of the
symmetric group HSP, measurements on a single register of the hidden subgroup
state reveal only an exponentially small amount of information about the
identity of the hidden subgroup.  In particular this means that if one makes
measurements which act only on a single hidden subgroup problem state at a
time, one cannot efficiently solve the HSP.  This was subsequently extended to
two registers of the hidden subgroup state by Moore and
Russell\cite{Moore:05b}, and then, in the culmination of this line of inquiry,
Hallgren {\em et al.}\cite{Hallgren:06a} showed that this extends all the way
up to the upper bound of Ettinger, Hoyer, and Knill.  In other words, for the
hidden subgroup problem relevant to the graph isomorphism problem, if one
attempts to efficiently solve this problem on a quantum computer one is
required to perform a measurement on $O(\log |{\mathcal G}|)$ registers
containing the hidden subgroup state in order to solve the problem.  In
particular if this measurement can be implemented, even adaptively, on less
than $k$ registers, then this measurement will not be able to solve the HSP.

The results of Hallgren {\em et al.} imply that in order to solve the HSP, one
must perform measurements on multiple copies of the hidden subgroup state in
order to efficiently solve the non-Abelian hidden subgroup problem.  One
important consequence of this is that the standard method combined with
performing a quantum Fourier transform on the hidden subgroup state, a paradigm
which works for many of the efficient quantum algorithms for the hidden
subgroup problem, cannot be used to find an efficient quantum algorithm for the
non-Abelian hidden subgroup problem.  In particular the above discussion makes
it clear that if there is hope for an efficient quantum algorithm for the
non-Abelian hidden subgroup problem in the standard paradigm, then measurements
across multiple copies of the hidden subgroup state must be used.  So far only
a small number of quantum algorithms have used such measurements.  The first
such algorithm for HSPs was in Kuperberg's subexponential time algorithm for
the dihedral hidden subgroup problem\cite{Kuperberg:03a}.  Recently Bacon,
Childs, and van Dam\cite{Bacon:05a} showed that for the Heisenberg group a
measurement across two hidden subgroup states could be used to efficiently
solve this HSP (and a measurement across $r$ hidden subgroup states could be
used to solve the hidden subgroup problem over ${\mathbb Z}_p^r \ltimes
{\mathbb Z}_p$.)   This result gives the first indication that while the
results of Hallgren {\em et al.} put a damper on traditional attempts to solve
the HSP using only the standard approach and the quantum Fourier transform, all
is not lost, and if there is any hope for efficient quantum algorithm for the
full non-Abelian HSP, techniques for efficient quantum measurements across
multiple copies of the hidden subgroup state must be developed.

The efficient algorithm for the Heisenberg hidden subgroup problem was
discovered by identifying an {\em optimal measurement} for the hidden subgroup
problem.  In the optimal measurement approach to the hidden subgroup problem,
one assumes that one has been given $k$ copies of the hidden subgroup state
$\rho_{\mathcal H}$ with an a priori probability $p_{\mathcal H}$.  One then
wishes to find the generalized measurement (which we hereafter refer to only as
a measurement) on these $k$ copies which maximizes the probability of
successfully identifying the hidden subgroup ${\mathcal H}$, averaged over the
a priori probabilities $p_{\mathcal H}$.  Cast in this form, the hidden
subgroup problem becomes a problem of optimal state discrimination.  A set of
necessary and sufficient conditions for a measurement to be such an optimal
measurement was discovered over thirty years ago by Holevo\cite{Holevo:73a} and
Yuen, Kennedy, and Lax\cite{Yuen:75a}.

Ip\cite{Ip:03a} was the first to consider the optimal measurement for the HSP.
In particular he examined the optimal measurement for the HSP when all of the
subgroups are given with equal a prior probability.  Ip showed that in this
case, for the Abelian HSP, the standard approach to solving the HSP is optimal.
Further Ip showed that for the dihedral hidden subgroup problem, the optimal
measurement was not to perform a quantum Fourier transform over the dihedral
group followed by a projective measurement.  Continuing on in this line of
inquiry, Bacon, Childs, and van Dam derived an exact expression for the optimal
measurement for the dihedral hidden subgroup problem\cite{Bacon:06c}.  This
measurement turned out to be the so-called pretty good
measurement\cite{Hausladen:94a}.  Further the optimal measurement on $k$ copies
of the hidden subgroup state was discovered to be a nontrivial measurement
across multiple copies of the hidden subgroup states.  Thus, even though
measurement on a single register containing the hidden subgroup state is enough
to information theoretically reconstruct the hidden subgroup state, a
measurement across many registers containing the hidden subgroup is optimal for
solving this problem.  Unfortunately, it is not known how to efficiently
implement the optimal measurement described in \cite{Bacon:06c}.  However,
building upon the optimal measurement approach, Bacon, Childs, and van Dam then
applied the apparatus of optimal measurements for the HSP to the HSP for
certain semidirect product groups of the for ${\mathbb Z}_p^k \rtimes {\mathbb
Z}$, for a fixed $k$ and prime $p$\cite{Bacon:05a}.  Again it was discovered
that the optimal measurement on multiple registers containing the hidden
subgroup states required a measurement over multiple registers containing the
hidden subgroup state (in fact over $r$ registers.)  However this time the
authors were able to find an efficient quantum algorithm implementing this
measurement.  Thus a non-trivial quantum algorithm for a non-Abelian hidden
subgroup problem was discovered which was optimally solved by a measurement
across multiple copies of the hidden subgroup state.

However, in spite of this success, there is much that remains mysterious about
the efficient quantum algorithm for the Heisenberg HSP.  In particular, why is
there an efficient algorithm for implementing the optimal measurement in this
case?  Is there any {\em structure} behind measurements across multiple copies
of hidden subgroup states which can be used to solve the hidden subgroup
problem?  In this paper we present partial answers to these questions and
highlight the role of an important transform over many registers containing the hidden subgroup state, the Clebsch-Gordan transform, in providing an efficient
algorithm.  We believe that this is an important insight, first of all because
it gives a new explanation for the efficient pretty good measurement based
algorithm of Bacon, Childs, and van Dam.  Further we believe that our result highlights the important role of that Clebsch-Gordan transform can play in quantum algorithms.  Clebsch-Gordan transforms, like quantum Fourier transforms over finite groups, suffer from there not being a canonical choice for the bases of registers output by these transforms.  In this work we show that by a judicious choice of this arbitrary basis, we can use Clebsch-Gordan transforms to solve a non-Abelian HSP.  Thus beyond showing that Clebsch-Gordan transforms can be useful for solving HSPs we can focus the search for efficient HSP algorithm to an even smaller problem of understanding the choose of basis for these transforms.

\section{The Hidden Subgroup Conjugacy Problem} \label{sec:hscp}

In this section we present a variant of the hidden subgroup problem which we
label the hidden subgroup conjugacy problem.

Two subgroups ${\mathcal H}_1 \subset {\mathcal G}$ and ${\mathcal H}_2
\subset{\mathcal G}$ are said to be conjugate to each other if there exists an
element of $g \in {\mathcal G}$ such that
\begin{equation}
{\mathcal H}_1=\{g h_2 g^{-1}, \forall h_2
\in {\mathcal H}_2 \}.
\end{equation}The notion of conjugate subgroups forms an equivalence
relationship.  Therefore we can classify all subgroups into distinct sets of
subgroups which are all conjugate to each other. The hidden
subgroup conjugacy problem (HSCP) is exactly like the HSP, but
instead of requiring that we correctly identify the hidden subgroup ${\mathcal
H}$ of a function, we only require than one correctly identify which set of
conjugate subgroups ${\mathcal H}$ belongs to.

Clearly solving the HSP in its original form will allow one
to solve the hidden subgroup conjugacy problem, but less is known about the
reverse relation.  First it is clear that when the group is Abelian or when the
subgroups are normal, the HSP is equivalent to the HSCP (since subgroups of
Abelian groups and normal subgroups are conjugate only to themselves.)
Recently Fenner and Zhang\cite{Fenner:06a} have examined the difference between
the search and decision version of the HSP (in the search problem one is
required to return the hidden subgroup and in the decision subgroup one is
required to distinguish whether the hidden subgroup is trivial or not.)  Their
results imply that the HSCP and the HSP over permutation groups are polynomial
time equivalent.  Similarly they show that for the dihedral group, when the
order of the group is the product of many small primes, then the HSCP and the
HSP are polynomial time equivalent.  In Section~\ref{sec:heis} we will show
that the HSCP and HSP are quantum polynomial time equivalent for the Heisenberg
group.

Finally we note that the HSP can be decomposed into the HSCP along with what we
call the hidden conjugate subgroup problem (HCSP).  In the hidden conjugate
subgroup, one is given a function which hides one of a set of subgroups all of
which are conjugate to each other and one desires to identify the conjugate
subgroup.  For the single copy HCSP, the pretty good measurement was shown to
be optimal by Moore and Russell\cite{Moore:05c}.

\section{Symmetry Considerations and the HSP} \label{sec:symhsp}

The hidden subgroup state of Eq.~(\ref{eq:hsstate}) possess a set of
symmetries
which allow us to, without loss of generality, perform a change of basis which
exploits these symmetries.  These symmetries are related to the regular
representations of the group.

\subsection{Regular Representations}

There are two regular representations of the group ${\mathcal G}$ which we
will
be interested in, the left regular representation and the right regular
representation.  Both of these representations act on a Hilbert space with a
basis labeled by the elements of the group ${\mathcal G}$.  Define the left
regular representation via its action on basis states of this Hilbert space,
\begin{equation}
R_L(g)|g^\prime\rangle =|g g^\prime\rangle,
\end{equation}
where $g g\prime$ is the element of ${\mathcal G}$ obtained by multiplying $g$
and $g^\prime$.  Similarly, define the right regular representation via
\begin{equation}
R_R(g)|g^\prime\rangle=|g^\prime g^{-1}\rangle.
\end{equation}
The regular representations are, in general, reducible representations of
${\mathcal G}$.  In fact these representations are particularly important in
the representation theory of finite groups.  The reason for this is that the
regular representations are reducible into a direct sum of all irreducible
representations (irreps) of the group ${\mathcal G}$.  Thus, it is possible to
find a basis in which $R_L$ acts as
\begin{equation}
R_L(g) = \bigoplus_{\mu} I_{d_\mu} \otimes D_\mu(g),
\end{equation}
where the direct sum is over all irreps of the group ${\mathcal G}$, $D_\mu(g)$
is the $\mu$th irrep
evaluated at group element $g$ and $d_\mu$ is the dimension of the irrep
$\mu$.
A similar decomposition occurs for the right regular representation,
\begin{equation}
R_R(g) = \bigoplus_{\mu} D_\mu(g) \otimes I_{d_\mu}.
\end{equation}
In fact, an elementary result of finite group representation theory tells us
that
the basis in which $R_L(g)$ acts as the above direct sum is also the basis in
which $R_R(g)$ acts as the above direct sum.  In other words, in this basis,
\begin{equation}
R_L(g) R_R(g^\prime)= R_R(g^\prime)R_L(g)=\bigoplus_\mu D_\mu (g^\prime)
\otimes D_\mu(g). \label{eq:regred}
\end{equation}
Let us call the basis described above the $|\mu,r_\mu,l_\mu\rangle$ basis
where
$r_\mu=1,\dots,d_\mu$ and $l_\mu=1,\dots,d_\mu$.

\subsection{Symmetry of Hidden Subgroup States}

Why are the left and right regular representations relevant to the hidden
subgroup problem?  Well it is easy to check that the hidden subgroup state
$\rho_{\mathcal H}$, for all hidden subgroups ${\mathcal H}$, are invariant
under conjugation by the left regular representation,
\begin{equation}
D_L(g) \rho_{\mathcal H} D_L(g^{-1}) = \rho_{\mathcal H}, \label{eq:sym}
\end{equation}
for all $g \in {\mathcal G}$.  The reason for this is that left multiplication
by a fixed group element acts as a permutation on left cosets.  Notice that
$\rho_{\mathcal H}$ is not invariant under a similar transform using the right
regular representation unless the representation being used is an element of
the subgroup ${\mathcal H}$.

What is the consequence of the invariance of the ${\rho}_{\mathcal H}$ with
respect to the left regular representation?  Schur's lemma\cite{Cornwell:97a}
tells us that if an
operator is invariant with respect to the operators enacting a representation
of the group, then that operator has support only over the commutant of the
this representation.  For our purposes the commutant is simply the algebra of matrices which commute with the left regular representation, $M D_L(g)=D_L(g) M$.  Since $\rho_{\mathcal H}$ is invariant with respect to
the
operators enacting the left regular representation of ${\mathcal G}$ and the
commutant of the left regular representation of ${\mathcal G}$ is the right
regular representation of ${\mathcal G}$, this means that in the
$|\mu,r_\mu,l_\mu\rangle$ basis, $\rho_{\mathcal H}$ can be expressed as
\begin{equation}
{\rho}_{\mathcal H} = \bigoplus_\mu \sigma_{\mu,{\mathcal H}} \otimes
I_{d_\mu},
\end{equation}
where $\sigma_{\mu,{\mathcal H}}$ is an operator with support only on the
space
acted upon by the irreducible representation of the right regular
representation.

What does this mean for obtaining information about ${\mathcal H}$ by making
measurements on ${\rho}_{\mathcal H}$?  For now we focus on a single
measurement on a single copy of ${\rho}_{\mathcal H}$.  Recall that a
generalized measurement is described by a set of positive operators
$\{{P}_1,\dots,{P}_k\}$ which sum to identity $\sum_{\alpha=1}^k {P}_\alpha =
{I}$. Outcomes of the measurement correspond to the indices, with the
probability of getting outcome $\alpha$ when measuring state ${\rho}$ given by
$p_\alpha={\rm Tr} \left[ {\rho} {P}_\alpha \right]$.  So if we make a
measurement with operator ${P}_\alpha$ on ${\rho}_{\mathcal H}$ the
probability
of getting outcome $\alpha$ is
\begin{equation}
p_\alpha= {\rm Tr} \left[ {\rho}_{\mathcal H} {P}_\alpha \right].
\end{equation}
But, since ${\rho}_H$ is invariant under the left regular representation
operators, we can express this probability as
\begin{equation}
p_\alpha = {\rm Tr} \left[{1 \over |{\mathcal G}|} \sum_{g \in {\mathcal G}}
{R}_L(g) {\rho}_{\mathcal H} { R}_L^\dagger(g) { P}_\alpha \right]= {\rm Tr}
\left[{\rho}_{\mathcal H} \tilde{P}_\alpha \right],
\end{equation}
where
\begin{equation}
\tilde{ P}_\alpha = {1 \over |{\mathcal G}|} \sum_{g \in {\mathcal G}} {
R}_L^\dagger(g) {P}_\alpha {R}_L(g).
\end{equation}
But $\tilde{P}_\alpha$ is ${P}_\alpha$ symmetrized over left regular
representation of the group ${\mathcal G}$.  This implies that
$\tilde{P}_\alpha$ commutes with the left regular representation,
\begin{equation}
R_L(g^\prime)^\dagger  \tilde{P}_\alpha R_L^\dagger(g^\prime)=
\tilde{P}_\alpha,
\end{equation}
for all $g$ in ${\mathcal G}$.  Thus using Schur's lemma, $\tilde{P}_\alpha$
has support over only the commutant
of the left regular representation, our good friend the right regular
representation.  In other words, in the $|\mu,r_\mu,l_mu\rangle$ basis, we find
that
\begin{equation}
\tilde{P}_\alpha= \bigoplus_\mu {Q}_{\mu,\alpha} \otimes {I}_{d_\mu}.
\end{equation}
This means that given some measurement ${P}_\alpha$, the probabilities of the
different outcomes $\alpha$ depends only on the symmetrized version of ${
P}_\alpha$.  Therefore without loss of generality we can deal with measurement
which are already symmetrized and therefore have the above decomposition over
the regular representation decomposition.

As a final consequence of the symmetry of the ${\rho}_{\mathcal H}$, we recall
that the algebra formed by a representation of a group (the group algebra) is
a
complete basis for operators which have support on the space the irreducible
representations of the representation of the group act.  For the right regular
representation, where we have
\begin{equation}
{R}_R(g) = \bigoplus_\mu {D}_{\mu}(g) \otimes {I}_{d_\mu},
\end{equation}
then if we look at linear combinations of the operators ${R}_R(g)$ for all $g
\in {\mathcal G}$, we find that these operators span the space of operators
defined by
\begin{equation}
\bigoplus_\mu {M}_{\mu} \otimes {I}_{d_\mu},
\end{equation}
for all choices of ${M}_\mu$.  Because ${\rho}_{\mathcal H}$ has support over
the irreducible representations of the right regular representations of
${\mathcal G}$ this means that
\begin{equation}
{\rho}_{\mathcal H}= \sum_{g \in {\mathcal G}} c_g({\mathcal H}) {R}_R(g).
\end{equation}
Notice that the dependence of the subgroup ${\mathcal H}$ is only in the
coefficients of the expansion.

Furthermore we can directly evaluate these coefficients in the expansion over
the right regular representation.  Recall that for the right regular
representation,
\begin{equation}
{\rm Tr} \left[ {R}_R(g) \right] = |{\mathcal G}| \delta_{g,e},
\end{equation}
where $e$ is the identity element of the group.  This implies
\begin{equation}
{\rm Tr} \left[ {R}_R(g) {R}_R^\dagger(g^\prime) \right] = {\rm Tr} \left[
{R}_R ((g^\prime)^{-1} g) \right] = |{\mathcal G}| \delta_{g,g^\prime}.
\end{equation}
Hence we find that
\begin{equation}
c_g({\mathcal H}) = {1 \over |{\mathcal G}|} {\rm Tr} \left[ {\rho}_{\mathcal
H} {R}_R(g) \right].
\end{equation}
Next we get tricky and note that
\begin{eqnarray}
{\rm Tr} \left[ {\rho}_{\mathcal H} {R}_R(g) \right] &=& {\rm Tr} \left[
{|{\mathcal H}| \over |{\mathcal G}|} \sum_{g^\prime={\rm coset~rep.}}
|g^\prime{\mathcal H} \rangle \langle g^\prime{\mathcal H}| {R}_R(g)
 \right]
 = {|{\mathcal H}| \over |{\mathcal G}|} \sum_{g^\prime={\rm
 coset~rep.}}\langle g^\prime {\mathcal H} | {R}_R(g) | g^\prime
 {\mathcal H} \rangle \nonumber \\
 &=& {|{\mathcal H}| \over |{\mathcal G}|} \sum_{g^\prime={\rm
 coset~rep.}}\langle g^\prime {\mathcal H}  | g^\prime
 {\mathcal H} g^{-1}\rangle,
\end{eqnarray}
where
\begin{equation}
|g^\prime{\mathcal H}g^{-1}\rangle = {1 \over \sqrt{|{\mathcal H}|}} \sum_{h
\in {\mathcal H}} |g^\prime hg^{-1}\rangle.
\end{equation}
But
\begin{eqnarray}
\langle g^\prime {\mathcal H}  | g^\prime
 {\mathcal H} g\rangle= {1 \over |{\mathcal H}|} \sum_{h_1,h_2 \in {\mathcal
 H}
 } \langle g^\prime h_1  | g^\prime h_2 g \rangle = {1 \over |{\mathcal H}|}
 \sum_{h_1,h_2 \in {\mathcal H}
 } \langle h_1  | h_2 g \rangle = \delta_{g,{\mathcal H}},
\end{eqnarray}
where $\delta_{g,{\mathcal H}}$ is shorthand for $1$ if $g \in {\mathcal H}$
and $0$ otherwise.  Therefore
\begin{equation}
c_g({\mathcal H}) = {1 \over |{\mathcal G}|} {\rm Tr} \left[ {\rho}_{\mathcal
H} {R}_R(g) \right] ={1 \over |{\mathcal G}|} \delta_{g,{\mathcal H}}.
\end{equation}
In terms of ${\rho}_{\mathcal H}$ this implies the neat little expression
\begin{equation}
{\rho}_{\mathcal H}= {1 \over |{\mathcal G}|} \sum_{h \in {\mathcal H} } {
R}_R(h). \label{eq:neat}
\end{equation}
For example, using the above expression, it is easy to check that
$\rho_{\mathcal H}$ is a proportional to a projector,
\begin{equation}
\rho_{\mathcal H}^2={1 \over |{\mathcal G}|^2} \sum_{h_1,h_2 \in {\mathcal H}}
{R}_R(h_1h_2) ={|{\mathcal H}| \over |{\mathcal G}|^2} \sum_{h \in {\mathcal
H}}R_R(h)={ |{\mathcal G}| \over |{\mathcal H}|} \rho_{\mathcal H}.
\end{equation}
Finally it is useful to write $\rho_{\mathcal H}$ in $|\mu,r_\mu,l_\mu\rangle$
basis
\begin{equation}
{\rho}_{\mathcal H}= \bigoplus_\mu \left[{1 \over |{\mathcal G}|} \sum_{h \in
{\mathcal H} } D_\mu(h)\right] \otimes I_{d_{\mu}}. \label{eq:neato}
\end{equation}

To recap this subsection, we have shown that the hidden subgroup state is
invariant under conjugation by the left regular representation.  This implies
that the state is block diagonal in a basis where the right and left regular
representations are fully reduced.  Finally this leads to an expression for the
hidden subgroup state in terms of the right regular representation,
Eq.~(\ref{eq:neat}), which in turn leads to a simple block-diagonal expression
for the hidden subgroup state, Eq.~(\ref{eq:neato}).

\subsection{The Quantum Fourier Transform over a Finite Group}

Having identified the symmetries of a hidden subgroup state $\rho_{\mathcal
H}$
and shown that this leads to $\rho_{\mathcal H}$ being block diagonal in a
particular basis, an important question is whether one can actually perform this
basis change efficiently on a quantum computer.  Indeed we haven't even
identified what this change of basis is.  In fact, the basis change is nothing
more than the
quantum fourier transform over the group ${\mathcal G}$.  The quantum fourier
transform (QFT) over the group ${\mathcal G}$ is defined as the unitary
transform
\begin{equation}
Q_{\mathcal G}=\sum_{g \in {\mathcal G}} \sum_\mu \sum_{i,j=1}^{d_\mu}
\sqrt{d_\mu \over |{\mathcal G}|}[D_{\mu}(g)]_{i,j} |\mu,i,j\rangle \langle
g|.
\end{equation}
It is then easy to check, using the orthogonality relationships of irreducible
representations that
\begin{eqnarray}
Q_{\mathcal G} R_L(g) Q_{\mathcal G}^\dagger&=& \bigoplus_\mu
\sum_{i=1}^{d_\mu} |i\rangle \langle i| \otimes \sum_{j,j^\prime=1}^{d_\mu}
[D_\mu]_{j,j^\prime}[g] |j\rangle \langle j^\prime|
\nonumber \\
Q_{\mathcal G} R_R(g) Q_{\mathcal G}^\dagger&=& \bigoplus_\mu
\sum_{j,j^\prime=1}^{d_\mu}[D_\mu]_{j,j^\prime}[g] |j\rangle \langle j^\prime|
\otimes \sum_{i=1}^{d_\mu}
|i\rangle \langle i|. \nonumber \\
\end{eqnarray}
Or in other words the QFT performs exactly the change of basis which block
diagonalizes the left and right regular representations into a irreducible
irreps as described in Eqs.~(\ref{eq:regred}).

When can the QFT over a group ${\mathcal G}$ be enacted by a quantum circuit of
size polynomial in $\log |{\mathcal G}|$?  While the full answer to this
question is not known, for many groups, including the important symmetric
group\cite{Beals:97a} and dihedral group\cite{Hoyer:97a,Puschel:99a}, efficient
quantum circuits for the QFT are known.  A quite general method for performing
efficient QFTs over a large class of finite groups, including the Heisenberg
group, has been derived by Moore, Rockmore, and Russell\cite{Moore:04b}.  We
refer the reader to the latter paper for more details on the quantum Fourier
transform.

Finally we should note that the QFT is a change of basis which is defined only
up to the choice of the basis for the $D_\mu$ irreps.  For our purposes, this
basis choice will not matter, and we refer the reader to the paper of Moore,
Rockmore, and Russell\cite{Moore:04b} for details on choices of this basis for
different groups.

\subsection{The Symmetry Exploiting Protocol} \label{sec:symprot}

We have now seen that the QFT over the group ${\mathcal G}$ is the transform
which block diagonalizes all hidden subgroup states.  By the symmetry argument
above we can, without loss of generality, apply this transform.  Further, since
$\rho_{\mathcal H}$ is an incoherent sum over the different irreps $\mu$, we
can, without a loss of generality, perform this transform, and then measure the
irrep index $|\mu\rangle$.  Further, since $\rho_{\mathcal H}$ acts trivially
over the space where the left regular representation acts, the $|l_\mu\rangle$
index contains no information about the hidden subgroup and can also be
measured (resulting in a uniformly random number between $1$ and $d_\mu$.)

Hence we see that we can recast the single copy hidden subgroup problem as
\begin{enumerate}
  \item Perform a QFT on the hidden subgroup state.
  \item Measure the irrep label $\mu$ and throw away the register where the
  left regular representation irrep acts.  The probability of obtaining irrep
  $\mu$ is given by
  \begin{equation}
  p_\mu[{\mathcal H}]={d_\mu \over |{\mathcal G}|} \sum_{h \in {\mathcal H}}
  \chi_\mu (h),
  \end{equation}
  where $\chi_\mu(g)={\rm Tr}[D_\mu(g)]$ is the character of the group element
  $g$ in the irrep $\mu$.
  \item One is then left with a state with support over only the space where
  the right regular representation irrep acts.  This state is given by
  \begin{equation}
  \rho_{\mu}[{\mathcal H}]={d_\mu \over |{\mathcal G}| p_\mu} \sum_{h \in
  {\mathcal H}} D_\mu(h),
  \end{equation}
  or, in other words the state
  \begin{equation}
  \rho_{\mu}[{\mathcal H}]={\sum_{h \in {\mathcal H}} D_\mu(h) \over
  \sum_{h^\prime \in {\mathcal H}} \chi_\mu(h^\prime)},
  \end{equation}
  assuming that $p_\mu \neq 0$.
\end{enumerate}
The above protocol has been designed based solely on the symmetry arguments of
the hidden subgroup state.  Notice that there are two locations already where
information can appear about the hidden subgroup.  The first is in the
probabilities $p_\mu[{\mathcal H}]$.  The second is in the state
$\rho_{\mu}[{\mathcal H}]$.

Finally note that if the hidden subgroup is the trivial group ${\mathcal
T}=\{e\}$, then we obtain the $\mu$th irrep with probability
\begin{equation}
p_\mu[{\mathcal H}]={d_\mu^2 \over |{\mathcal G}|}. \label{eq:trivprob}
\end{equation}
Note as a sanity check that this is indeed a probability since the sum of the
squares of the dimensions of all irreps is the order of the group.

\subsection{Multiple Copies of the Hidden Subgroup State}

As we have discussed in the introduction to the HSP, we are
most interested in the setting where we use multiple copies of a hidden
subgroup state to determine the hidden subgroup.  In this setting, each of the
hidden subgroup states will retain the symmetry we have described above.  In
other words our state is $\rho_{\mathcal H}^{\otimes m}$ and each of these can
be reduced into irreducible irreps as described in Eq.~(\ref{eq:neato}).  In
this case we immediately note that there is another symmetry of these multiple
copies of the same state.  This symmetry is a permutation symmetry.  If we
permute the different copies, $\rho_{\mathcal H}^{\otimes m}$ is invariant. In
this paper we will not explore this symmetry, only noting that measuring the
irrep label alone for the Schur transform cannot be used to solve either the
HSP or the HSCP.  For a negative result in using permutation symmetries in
trying to solve the HSP, we point the reader to the work of Childs, Harrow, and
Wocjan\cite{Childs:06a}.

\section{Symmetry Considerations and the HSCP} \label{sec:symhscp}

We have seen in the previous section that the symmetry of the hidden subgroup
states means that these states have a structure related to the left and right
regular representations of the group being considered.  This led, in turn, to
the idea that one should exploit this structure by performing a quantum
Fourier
transform over the group ${\mathcal G}$ on the hidden subgroup states.  In this
section we will turn our
attention to the hidden subgroup conjugacy problem and show how similar
arguments lead one to a different transform than the QFT, the Clebsch-Gordan
transform over the group ${\mathcal G}$.

\subsection{Single Copy Case}

Recall that in the HSCP we wish to determine which set of conjugate subgroups
a
hidden subgroup state belongs to.  In particular we wish to design an
algorithm
which learns the set of conjugate subgroups hidden by $f$, but not the
particular hidden subgroup.  Stated in this manner, it is then natural to
define the hidden subgroup conjugacy states which are an incoherent sum of all
of the hidden subgroup states which belong to a given set of conjugate
subgroups.  Thus we can define the single copy hidden subgroup conjugacy state
as
\begin{equation}
\rho_{[{\mathcal H}]}={1 \over |{\mathcal G}|} \sum_{g \in {\mathcal G}}
\rho_{g{\mathcal H}g^{-1}},
\end{equation}
where $g {\mathcal H}g^{-1}=\{ghg^{-1},h \in {\mathcal H}\}$.  Notice that we
have summed over the entire group here and not over simply the different
conjugate subgroups.  However, since every conjugate subgroup will appear an
equal number of times in this sum, and we have normalized by the size of the
group, our state is equivalent to a state which is one of the elements of the
set of conjugate subgroups with equal probability.

Using Eq.~(\ref{eq:neat}) it is easy to see that
\begin{eqnarray}
\rho_{g{\mathcal H}g^{-1}}={1 \over |{\mathcal G}|} \sum_{h \in {\mathcal H}}
R_R(g h g^{-1})
={1 \over |{\mathcal G}|} \sum_{h \in {\mathcal H}}R_R(g) R_R(h)R_R(g^{-1}),
\end{eqnarray}
so that
\begin{equation}
\rho_{[{\mathcal H}]}={1 \over |{\mathcal G}|} \sum_{g \in {\mathcal G}}
R_R(g)
\rho_{\mathcal H} R_R(g^{-1}).
\end{equation}
In other words, $\rho_{[{\mathcal H}]}$ is $\rho_{\mathcal H}$ symmetrized
over
the right regular representation.  This implies that
$\rho_{[{\mathcal H}]}$ is invariant under the right regular representation:
\begin{equation}
R_R(g) \rho_{[{\mathcal H}]}R_R(g^{-1})=\rho_{[{\mathcal H}]}.
\end{equation}
Thus we can again apply Schur's lemma.  Since $\rho_{\mathcal H}$ already only
has
support over the right regular representation, we can deduce that
$\rho_{[{\mathcal H}]}$ in the $|\mu,r_\mu,l_\mu\rangle$ basis is given by
\begin{equation}
\rho_{[{\mathcal H}]}= \bigoplus_\mu c_\mu({\mathcal H}) I_{d_\mu} \otimes
I_{d_\mu}. \label{eq:weak}
\end{equation}
We can use the character projection operator, which projects onto a given
irrep
space,
\begin{equation}
C_\mu={d_\mu \over |{\mathcal G}|} \sum_{g \in {\mathcal G}} \chi_\mu(g)^*
R_R(g),
\end{equation}
where $\chi_\mu(g)$ is the character of the $\mu$th irrep evaluated at group
element $g$, along with our simple representation of the hidden subgroup
states
to determine $c_\mu({\mathcal H})$,
\begin{equation}
c_\mu({\mathcal H})={1 \over d_{\mu}^2} {\rm Tr} [C_\mu \rho_{[{\mathcal
H}]}]={1 \over |{\mathcal G}| d_\mu} \sum_{h \in {\mathcal H} } \chi_\mu(h)^*.
\end{equation}

What does our expression for the hidden subgroup conjugacy state in
Eq.~\ref{eq:weak} tell us?  It tells us that, without loss of generality, if we
are trying to solve the HSCP using a single copy of the hidden subgroup state,
then we can without loss of generality perform a QFT over ${\mathcal G}$ and
then measuring the irrep label $\mu$.  Further all of the information, if there
is enough information to reconstruct which set of conjugate subgroups the
hidden subgroup belongs to, can be derived from such a measurement.

\subsection{Multiple Copy Case} \label{sec:multihscp}

Above we have argued that $\rho_{[{\mathcal H}]}$ is the appropriate state to
consider when attempting to describe algorithms for the HSCP when we have been
given a single copy of the hidden subgroup state.  What state should we
consider for the multiple copy case?  At first glance one is tempted to answer
$\rho_{[{\mathcal H}]}^{\otimes m}$.  However, this is the state which is
relevant if we are attempting to identify the which set of conjugate subgroups
a hidden subgroup state belongs to, and we are given different hidden subgroup
states from this set every time we produce a hidden subgroup state.  This
however, is not our case, as we are still querying a function which hides a
fixed subgroup ${\mathcal H}$.  Thus the actual state which is relevant for
the
HSCP is instead
\begin{equation}
\rho_{[{\mathcal H}],m}={1 \over |{\mathcal G}|}\sum_{g \in {\mathcal G}}
\rho_{g {\mathcal H} g^{-1}}^{\otimes m}. \label{eq:hscsmulti}
\end{equation}

The nice thing about Eq.~(\ref{eq:hscsmulti}) is that it represents a state
which has been averaged over the $m$-fold direct product of the right regular
representation. The $m$-fold direct product of right regular representations is
the representation of ${\mathcal G}$ given by $R_R(g)^{\otimes m}$, i.e. the
tensor product of the right regular representation acting in the same manner on
every tensor product space.  Indeed we can express Eq.~(\ref{eq:hscsmulti}) as
\begin{equation}
\rho_{[{\mathcal H}],m}={1 \over |{\mathcal G}|}\sum_{g \in {\mathcal G}}
R_R(g)^{\otimes m} \rho_{\mathcal H}^{\otimes m} R_R(g^{-1})^{\otimes m}.
\end{equation}
This implies, as before, that $\rho_{[{\mathcal H}],m}$ is invariant under the
$m$-fold direct product of the right regular representation,
\begin{equation}
R_R(g)^{\otimes m} \rho_{[{\mathcal H}],m} R_R(g^{-1})^{\otimes
m}=\rho_{[{\mathcal H}],m}. \label{eq:hscpsym}
\end{equation}
This is the symmetry which we are going to exploit to help us solve the
multicopy HSCP.

In order to exploit the symmetry in Eq.~(\ref{eq:hscpsym}) we begin by first
rewriting $\rho_{[{\mathcal H}]}$.  In particular we rewrite $\rho_{[{\mathcal
H}]}$ in a basis in which every hidden subgroup state copy has been subject to
a QFT over ${\mathcal G}$:
\begin{equation}
\rho_{[{\mathcal H}],m}={1 \over |{\mathcal G}|^{k+1}}\sum_{g \in {\mathcal G}}
\bigotimes_{i=1}^m \left[ \bigoplus_{\mu_i}\sum_{h_i \in {\mathcal H}}
D_{\mu_i}(g h_i g^{-1})  \otimes I_{d_{\mu_i}}  \right].
\end{equation}
Exchanging the product of sums with a sum of products we can express this as
\begin{equation}
{1 \over |{\mathcal G}|^{k+1}}\sum_{g \in {\mathcal G}}
\bigoplus_{\mu_1,\dots,\mu_m}
\bigotimes_{i=1}^m \left[\sum_{h_i \in {\mathcal H}} D_{\mu_i}(g h_i g^{-1})
\otimes I_{d_{\mu_i}}\right].
\end{equation}
From this expression we see that we should focus on terms for a fixed choice of
single copy irrep labels, $\mu_1,\dots,\mu_m$.
\begin{equation}
M={1 \over |{\mathcal G}|^{k+1}} \sum_{g \in {\mathcal G}} \bigotimes_{i=1}^m
\sum_{h \in {\mathcal H}}D_{\mu_i}(g h_i g^{-1}),
\end{equation}
which is invariant under the direct product action of the group,
\begin{equation}
\left[\bigotimes_{i=1}^m D_{\mu_i}(g)\right] M \left[\bigotimes_{i^\prime=1}^m
D_{\mu_{i^\prime}}(g^{-1})\right]=M.
\end{equation}
Thus we see that the symmetry of Eq.~(\ref{eq:hscsmulti}) is going to be
related to direct product representation $\bigotimes_{i=1}^m D_{\mu_i}(g)$.

What can we say about the direct product of $m$ irreps represented by
$\bigotimes_{i=1}^m D_{\mu_i}(g)$?  Well fist we note that this is, of course,
a representation of the group ${\mathcal G}$.  As such it will be, in general,
reducible.  In other words, there exists a basis under which
\begin{equation}
\bigotimes_{i=1}^m D_{\mu_i}(g)=\bigoplus_{\mu} I_{n_\mu} \otimes D_{\mu}(g).
\label{eq:cgdef}
\end{equation}
where $n_\mu$ is the multiplicity of the $\mu$th irrep in this decomposition.
We will call the transform which enacts the above basis change the $m$-fold
Clebsch-Gordan transform.  Notice that this transform can act in a
non-separable manner over the different spaces where each irrep lives.  It is
this transform which we will use below to solve the Heisenberg hidden subgroup
problem.

Given the basis change described in Eq.~(\ref{eq:cgdef}), and the fact that
$R_R(g)^{\otimes m}$ commutes with $\rho_{[{\mathcal H}],m}$, we can use
Schur's lemma to show that $\rho_{[{\mathcal H}],m}$ has a block diagonal form
related to this basis change.  In particular, the state $\rho_{[{\mathcal
H}],m}$ will act trivially on the space where the $\mu$ irrep acts in
Eq.~(\ref{eq:cgdef}) and will only have non-trivial support over the space
arising from the multiplicity of the $\mu$th irrep, i.e. the space where
identity acts in Eq.~(\ref{eq:cgdef}).

To be concrete let us define the $m$-fold Clebsch-Gordan transform.  It is the
transform which takes as input the $m$ irrep labels $|\mu_1\rangle \otimes
\cdots \otimes |\mu_m\rangle$ along with the spaces upon which these irreps act
$|v_i\rangle \otimes \cdots \otimes |v_m\rangle$ where $v_i=1,\dots,d_{\mu_i}$
and transforms this into the basis given by Eq.~(\ref{eq:cgdef}), which is
$|\mu_1\rangle \otimes \cdots \otimes |\mu_m\rangle \otimes |\mu\rangle \otimes
|w\rangle \otimes |v\rangle$, where $\mu$ is the irrep label for the subspaces
on the RHS of Eq.~(\ref{eq:cgdef}), $w=1,\dots,n_\mu$ labels the multiplicity
of these irreps, and $v=1,\dots,d_\mu$ labels the space where the $\mu$th irrep
acts.  Notice that we have not defined an explicit basis for $|v\rangle$ and
$|w\rangle$, just as in the QFT over a finite group, there was a choice in the
basis for the $|r_\mu\rangle$ and $|l_\mu\rangle$ basis.  When we encounter the
Clebsch-Gordan transform relevant for the Heisenberg HSCP, we will pick a
particular basis.

Given the above definitions we can now describe the a protocol for
distinguishing the hidden subgroup conjugacy states when we are given $m$
copies of the hidden subgroup state just as we did for the hidden subgroup
states:
\begin{enumerate}
  \item Perform QFTs over ${\mathcal G}$ over all $m$ hidden subgroup states.
  \item Measure the irrep label for each of these $m$ states, resulting in
  outcome $\mu_i$.
  \item Throw away the registers where the left regular representation irreps
  act.
  \item Take the $m$ remaining spaces where the right regular representation
  acts, along with the irrep labels, and perform a Clebsch-Gordan transform.
  \item Measure the irrep label after the Clebsch-Gordan transform, $\mu$.
  \item Throw away the space where the irrep acts nontrivially in the
  Clebsch-Gordan transform, leaving only the space arising from the
  multiplicity of the $\mu$th irrep.  From this remaining state, a measurement
  should be performed which solves the hidden subgroup conjugacy problem, along
  with the classical data resulting from the irrep measurements,
  $\mu_1,\dots,\mu_m$ and $\mu$.
\end{enumerate}
The above protocol can be used without a loss of computational power for
distinguishing hidden subgroup conjugacy states.  Notice that the final step is a step which requires a judicious choice of measurement.  Since the Clebsch-Gordan transform does not yield a canonical basis for the multiplicity space, this implies that all of the difficulty of distinguishing HSCP states lies in determining a basis for this measurement.  Further, getting into this basis should be accomplished using an efficient quantum circuit.  At this point one might wonder whether it is every possible to make such a judicious choice.  It is precisely the goal of this paper to show that such a choice is possible and can lead to efficient quantum algorithms for the HSCP.

\section{The Heisenberg Group} \label{sec:heis}

We now turn from our general discussion of the HSP and the HSCP for any finite
group ${\mathcal G}$ to a discussion where ${\mathcal G}$ is a particular
group, the Heisenberg group.  In this section we collect much of the relevant
information about this group, its subgroups, and its irreps.

We are interested in the Heisenberg group ${\mathcal H}_p={\mathbb Z}_p
\rtimes
{\mathbb Z}_p^2$ where $p$ is prime.  This is the group of upper right
triangular $3 \times 3$ matrices with multiplication and addition over the
field ${\mathbb F}_p$.  We denote elements of this group by a three-tuple of
numbers, $(x,y,z)$, with $x,y,z \in {\mathbb Z}_p$.  The multiplication rule
for this group is then given by
\begin{equation}
(x,y,z) (x^\prime,y^\prime,z^\prime)= (x+x^\prime, y+y^\prime +x
z^\prime,z+z^\prime).
\end{equation}
The inverse of an element $(x,y,z)$ of the Heisenberg group is the element
$(-x,-y+xz,-z)$.

The representation theory and subgroup structure of the Heisenberg group is
easily deduced.  We will use (up to a change in the way we write group
elements) the notation from the paper of Radhakrishnan, R{\"{o}}tteler and
Sen\cite{Radhakrishnan:05a}.  The Heisenberg group has $p^2$ different one
dimensional
irreps and $p-1$ different $p$ dimensional irreps.  The one dimensional irreps
are given
by
\begin{equation}
\chi_{a,b}((x,y,z))=\omega^{ax+bz},
\end{equation}
where $a,b \in {\mathbb Z}_p$ and $\omega=\exp\left( {2 \pi i\over p}
\right)$.
The $p$ dimensional irreps are given by
\begin{equation}
\sigma_k((x,y,z))= \omega^{ky} \sum_{r \in {\mathbb Z}_p} \omega^{kzr}
|r+x\rangle \langle r|,
\end{equation}
where $k \in {\mathbb Z}_p^*$.  For future reference, the characters of the $p$
dimensional irreps are
\begin{equation}
\chi_k((x,y,z))=\delta_{x,0}\delta_{z,0} \omega^{ky} p.
\end{equation}

The Heisenberg group has subgroups of four different orders, $p^3$, $p^2$,
$p$, and $1$. In Table~\ref{tab:subgroups} we catalog these subgroups and
describe their generators.  The first two groups in Table~\ref{tab:subgroups}
are the self-explanatory group itself and the trivial subgroup.  The ${\mathcal
N}_\alpha$
groups are normal subgroups of ${\mathcal H}_p$, ${\mathcal  N}_\alpha \lhd
{\mathcal
H}_p$.  These groups are isomorphic to the Abelian group ${\mathbb Z}_p \times
{\mathbb Z}_p$.  The ${\mathcal A}_{i,j}$ subgroups are Abelian subgroups which
are not
normal in ${\mathcal H}_p$. These groups are subgroups of the appropriate
${\mathcal N}_i$ subgroups.  In particular they are normal subgroups ${\mathcal
A}_{\alpha,j} \lhd
{\mathcal N}_\alpha$ and ${\mathcal A}_{\alpha,j} \nsubseteq {\mathcal
N}_{\alpha^\prime}$
if $i \neq i^\prime$.  Finally, ${\mathcal C}$ is the center of ${\mathcal
H}_p$.
\begin{table}[h]
\tcaption{Subgroups of the Heisenberg group, ${\mathcal H}_p$.}
\centerline{\footnotesize\smalllineskip
\begin{tabular}{|c|c|c|c|}
  \hline
  Subgroup & Generators & Label Range & Order \\
  \hline
  ${\mathcal H}_p$ & $(1,0,0),(0,1,0),(0,0,1)$ & -  & $p^3$  \\
  ${\mathcal T}$ & $(0,0,0)$ &  - & $1$ \\
  ${\mathcal N}_i$ & $(1,0,i),(0,1,0)$ &  $i \in {\mathbb Z}_p$ & $p^2$\\
  ${\mathcal N}_\infty$ & $(0,1,0),(0,0,1)$ & - & $p^2$ \\
  ${\mathcal A}_{i,j}$ & $(1,j,i)$ & $i,j \in {\mathbb Z}_p$ & $p$ \\
  ${\mathcal A}_{\infty,j}$ & $(0,j,1)$ & $ j \in {\mathbb Z}_p$ & $p$ \\
  ${\mathcal C}$ & $(0,1,0)$ & - & $p$ \\
  \hline
\end{tabular}}
\label{tab:subgroups}
\end{table}

We are interested in solving the HSP for ${\mathcal H}_p$.
A theorem from Bacon, Childs, and van Dam\cite{Bacon:05a} (which is an
extension of a theorem of Ettinger and Hoyer\cite{Ettinger:00a}) shows that
this problem can be reduced to the hidden
subgroup problem when the subgroup being hidden is restricted to being either
${\mathcal A}_{i,j}$, $i,j \in {\mathbb Z}_p$ or the trivial subgroup
${\mathcal T}$.  In fact we really only need to identify ${\mathcal A}_{i,j}$,
$i,j \in {\mathbb Z}_p$ since if our algorithm returns a subgroup of this form,
we can easily verify, by querying the function twice, whether it is constant on
left cosets this subgroup.  If the hidden subgroup were a trivial subgroup, we
would find the function is not constant and would then return that the hidden
subgroup is trivial.

We are interested in the hidden subgroup conjugacy problem.  Thus we would like
to know  which of the subgroups ${\mathcal A}_{i,j}$ are conjugate to each
other.  Conjugating a general element of ${\mathcal H}_p$ about a generator of
${\mathcal A}_{i,j}$, we find that
\begin{eqnarray}
(x,y,z) (1,j,i) (x,y,z)^{-1} =(1,j+xi-z,i).
\end{eqnarray}
Thus we see that subgroups ${\mathcal A}_{i,j}$ and ${\mathcal A}_{i,k}$ for
all $j,k \in {\mathbb Z}_p$ are conjugate to each other.  Thus in the HSCP we
are required to identify not $i$ and $j$ of $A_{i,j}$ (and distinguish this
from the trivial subgroup), but only to identify $i$.

We are now in a position to show that solving the HSCP for the Heisenberg group
can be used to solve the HSP for this group.  Suppose that we have solved the
HSCP for the Heisenberg group restricted to subgroups ${\mathcal A}_{i,j}$ and
we identified $i$.  We can identify $j$ via the observation that ${\mathcal
A}_{i,j}$ is normal in ${\mathcal N}_i$.  In other words, if we now work with a
HSP where our original $f$ is restricted to the subgroup ${\mathcal N}_i$
(which we explicitly know, since we have identified $i$) then we can use the
efficient quantum algorithm for finding normal subgroups\cite{Hallgren:00a} to
find $j$.  Alternatively (and equivalently, really) we could have simply noted that ${\mathcal N}_i$ is Abelian and thus we can run the standard HSP algorithm for Abelian groups over this to identify ${\mathcal A}_{i,j}$.  Thus we see that for the Heisenberg group, the HSP and the HSCP are actually quantum polynomial time equivalent to each other.

\subsection{The Heisenberg Hidden Subgroup State}

The hidden subgroup state for the Heisenberg group can be readily calculated
using the formalism described in the previous sections.  We are particularly
interested in a procedure where we perform a QFT over the Heisenberg group on
the hidden subgroup state, and measure the irrep label, as prescribed in the
protocol of Sec.~\ref{sec:symprot}.

We need to study the case that the hidden subgroup is one of the ${\mathcal
A}_{i,j}$ subgroups.  It is useful to note that this subgroup is the set
${\mathcal A}_{i,j}=\{ (l,2^{-1}l(l-1)i+lj,li),\forall l \in {\mathbb Z}_p\}$.
Thus the probability of observing the one dimensional irrep with character
$\chi_{a,b}$ is
\begin{equation}
p_{a,b}={1 \over p^3} \sum_{l \in {\mathbb Z}_p} \omega^{al+bli}={1 \over p^2}
\delta_{a,bi}.  \label{eq:oned}
\end{equation}
The probability of observing the $p$-dimensional irrep with character $\chi_k$
is
\begin{equation}
p_k={1 \over p^3} \sum_{l \in {\mathbb Z}_p} \chi_k((l,2^{-1}l(l-1)i+lj,li))={1
\over p}.
\end{equation}
Comparing this to the results from the trivial subgroup, we see that the
trivial and the ${\mathcal A}_{i,j}$ subgroups cannot be distinguished by the
above measurement of $\mu$.  Indeed, the probability of getting one of the $p$
dimensional irreps in both cases is uniformly ${1 \over p}$ and there are $p-1$
such irreps such that with probability $1-\frac{1}{p}$ the states are not
distinguished using just this irrep measurement.

We are interested in more than just the probability of the measurement of the
irrep label.  In particular we are most interested in the exact form of the
hidden subgroup state produced after we measure the irrep index and throw away
the index of the left regular representation.  When the hidden subgroup is
${\mathcal A}_{i,j}$ and we measure the one dimensional irrep with character
$\chi_{a,b}$ we obtain, of course, a single one dimensional space with density
matrix just the scalar $1$.  However when obtain one of the $p$ dimensional
irreps, we obtain the state over the $p$ dimensional irrep space,
\begin{equation}
\rho_k({\mathcal A}_{i,j})={1 \over p} \sum_{l \in {\mathbb Z}_p}
\omega^{2^{-1}l(l-1)i k+ljk} \sum_{r \in {\mathbb Z}_p} \omega^{kli
r}|r+l\rangle \langle r|. \label{eq:rhok}
\end{equation}
This latter state is the state which we will be most relevant to our efficient
algorithm for the HSCP over the Heisenberg group.

\section{The Clebsch-Gordan Transform over the Heisenberg Group}
\label{sec:cgheis}

If we have two irreps of the Heisenberg group, $D_{\mu_1}$ and $D_{\mu_2}$ what
are the rules for decomposing the direct product of these irreps into irreps?
In other words, what is the Clebsch-Gordan transform for the $2$-fold direct
product of two irreps of the Heisenberg group?

The first case to consider is when both of the irreps are one dimensional
irreps with characters $\chi_{a_1,b_1}$ and $\chi_{a_2,b_2}$.  In this case it
is easy to see that the direct product of these irreps is a one dimensional
irrep with corresponding character $\chi_{a_1+a_2,b_1+b_2}$ where the addition
is done over ${\mathbb Z}_p$.  The second case to consider is when one of the
irreps is the one dimensional irrep with character $\chi_{a_1,b_1}$ and the
second irrep is a $p$ dimensional irrep with character $\chi_{k_2}$.  In this
case it is easy to see that
\begin{eqnarray}
\chi_{a_1,b_1}((x,y,z)) \otimes \sigma_{k_2}((x,y,z))
= \omega^{k_2y+a_1x+b_1 z} \sum_{r \in {\mathbb Z}_p} \omega^{k_2zr}
|r+x\rangle \langle r|
\end{eqnarray}
Now making the change of basis described by the unitary matrix
\begin{equation}
V=\sum_{t \in {\mathbb Z}_p} |t+k_2^{-1}b_1\rangle \langle t|\left[\sum_{ s \in
{\mathbb Z}_p} \omega^{-a_1s} |s\rangle \langle s|\right],
\end{equation}
to $\chi_{a_1,b_1}((x,y,z)) \otimes \sigma_{k_2}((x,y,z))$ yields
\begin{equation}
V\omega^{k_2y+a_1x+b_2 z} \sum_{r \in {\mathbb Z}_p} \omega^{k_2zr}
|r+x\rangle \langle r|V^\dagger=\sigma_k((x,y,z)).
\end{equation}
Thus the direct product of the one dimensional irrep with character
$\chi_{a,b}$ and the $p$ dimensional irrep $\sigma_k$ is a $p$ dimensional
irrep $\sigma_k$.  Further there is no multiplicity in this decomposition.

The final case to consider is the case where one is forming the direct product
of two $p$ dimensional irreps,  $\sigma_{k_1}$ and $\sigma_{k_2}$.  In this
case we see that
\begin{equation}
\sigma_{k_1}((x,y,z)) \otimes \sigma_{k_2}((x,y,z))= \sum_{r_1 \in {\mathbb
Z}_p} \omega^{k_1 y +k_1 zr_1}
|r_1+x\rangle \langle r_1| \otimes \sum_{r_2 \in {\mathbb Z}_p} \omega^{k_2 y+
k_2 zr_2}
|r_2+x\rangle \langle r_2|.
\end{equation}
Consider first the case where $k_1+k_2 \neq 0 {\rm ~mod~}p$.  Define the
following change of basis
\begin{equation}
W=\sum_{a,b \in {\mathbb Z}_p} |a-b\rangle\langle a| \otimes |(k_1 a+ k_2
b)(k_1+k_2)^{-1} \rangle \langle b|. \label{eq:cgW}
\end{equation}
Applying this transform to $\sigma_{k_1}((x,y,z)) \otimes
\sigma_{k_2}((x,y,z))$ results in
\begin{equation}
 \sum_{r_1,r_2\in {\mathbb Z}_p}\omega^{k^\prime y +(k_1r_1+k_2 r_2)z}|r_1-r_2
 \rangle \langle r_1-r_2|
\otimes | (k_1r_2+k_2r_2)(k^\prime)^{-1}+x \rangle
\langle(k_1r_2+k_2r_2)(k^\prime)^{-1} |
\end{equation}
where $k^\prime=k_1+k_2$.  Using $u=r_1-r_2$ and $v=(k_1 r_1 +k_2
r_2)(k^\prime)^{-1}$ this can be reexpressed as
\begin{equation}
\omega^{k^\prime y} \sum_{u \in {\mathbb Z}_p} | u \rangle \langle u | \otimes
\sum_{v \in {\mathbb Z}_p }\omega^{k^\prime vz } |v+x\rangle \langle v|
\end{equation}
which we see is just $I \otimes \sigma_{k^\prime} ((x,y,z))$.  Thus when
$k_1+k_2 \neq 0{\rm ~mod~}p$, the direct product of the two $p$ dimensional
irreps labeled by $k_1$ and $k_2$ is reducible to the irrep labeled by
$k^\prime=k_1+k_2$ with multiplicity $p$.

When  $k_1+k_2 = 0{\rm ~mod~}p$ we find that
\begin{eqnarray}
\sigma_{k_1}((x,y,z)) \otimes \sigma_{k_2}((x,y,z))=\sum_{r_1,r_2 \in {\mathbb
Z}_p } \omega^{k_1z ( r_1 -r_2) } |r_1+x \rangle \langle r_1|
\otimes |r_2+x\rangle \langle r_2|.
\end{eqnarray}
Consider the unitary change of basis
\begin{equation}
X={1 \over \sqrt{p}}\sum_{a,b,c \in {\mathbb Z}_p} \omega^{(a+b)c}|a-b\rangle
\langle a| \otimes |c\rangle \langle b|.
\end{equation}
This transforms the direct product to
\begin{equation}
\sum_{u,c \in {\mathbb Z}_p} \omega^{k_1 uz+2xc} |u\rangle \langle u| \otimes
|c\rangle \langle c|,
\end{equation}
where $u=r_1-r_2$ and we have summed over $v=r_1+r_2$.  This we recognize as
$p^2$ one dimensional irreps, with every such irrep appearing exactly once.

Above we have derived how the direct product of two irreps of the Heisenberg
group decomposes into new irreps.  Can we efficiently enact a quantum algorithm
to achieve this transformation?  Certainly.  In fact we have done most of the
heavy lifting already in identifying the transforms $V$, $W$, and $X$.  Recall
the Clebsch-Gordan transform will act from a space with two irrep label
registers, $|\mu_1\rangle \otimes |\mu_2\rangle$, along with the space upon
which these irreps act $|v_1\rangle \otimes |v_2\rangle$ and have an output
space where the two irrep labels, $|\mu_1\rangle \otimes |\mu_2\rangle$ are
kept around and the direct product irrep label $|\mu\rangle$ is produced along
with the space where this irrep acts, $|v\rangle$ and the space of the
multiplicity of this irrep, $|w\rangle$.

The algorithm for efficiently enacting the Clebsch-Gordan then proceeds as
follows.  First, notice the transform above can all be done conditionally on
the $|\mu_1\rangle$ and $|\mu_2\rangle$ registers.  In other words, given that
we can efficiently enact the appropriate transform for fixed classical labels
of $|\mu_1\rangle$ and $|\mu_2\rangle$, then we can efficiently implement the
full Clebsch-Gordan transform by using the appropriate conditional gates.  Thus
we can divide up our algorithm into the cases we described above.

\begin{enumerate}
  \item ($\mu_1$ and $\mu_2$ are one dimensional irreps) In this case it is
  easy to efficiently classically compute the new irrep label from the old
  irrep labels, $a_1,b_1$ and $a_2,b_2$.  Indeed the new label is simply
  $a=a_1+a_2$ and $b=b_1+b_2$ done with addition modulo $p$.  Further the
  spaces involved are all one dimensional, ($|v_1\rangle$ and $|v_2\rangle$,
  along with $|v\rangle\otimes |w\rangle$)  Thus no other work needs to be done
  for this portion of the Clebsch-Gordan transform.
  \item ($\mu_1$ and $\mu_2$ are one and $p$ dimensional irreps)  In this case
  the new irrep label $\mu$ is nothing more than the original irrep label of
  the $p$ dimensional irrep, so such a label should be copied (reversibly
  added) into this register.  If we wish for the irreps to be expressed in the
  same basis that we express the $\sigma_k$ irreps, then we will need to apply
  the $V$ gate to the $|v_1\rangle \otimes |v_2\rangle$ register.  It is easy
  to see that $V$ can be enacted using classical reversible computation plus a
  diagonal phase gate which is easy to efficiently enact.
  \item ($\mu_1=k_1$ and $\mu_2=k_2$ are both $p$ dimensional irreps, such that
  $k_1+k_2 \neq 0~{\rm mod}~p$) In this case the new irrep label register will
  need to hold $k_1 + k_2 ~{\rm mod }~p$ which can easily be calculated using
  reversible classical circuits.  In addition, the transform $W$ must be
  enacted on the vector space $|v_1 \rangle \otimes |v_2\rangle$.  $W$ can also
  be enacted using classical reversible circuits efficiently.
\item ($\mu_1=k_1$ and $\mu_2=k_2$ are both $p$ dimensional irreps, such that
$k_1+k_2= 0~{\rm mod}~p$)  This is the only case where we cannot compute the
new irrep label directly.  Instead we must enact $X$, which can be done with a
combination of classical reversible circuits and an efficient QFT (or
approximation thereof) over ${\mathbb Z}_p$.  Once this is done, the new irrep
label can be found by simply transforming the vector basis labels $|u\rangle$
and $|c\rangle$ to the irrep label $|a=2u,b=k_1c\rangle$.
\end{enumerate}
Thus we have shown that the Clebsch-Gordan transform for the Heisenberg group
can be implemented with an efficient quantum circuit.

\subsection{Clebsch-Gordan Transform Methods}

For completeness we would here like to discuss prior results for obtaining
transforms very similar to Clebsch-Gordan transforms over finite groups, and
why it is important to distinguish these methods from the method we have
described above.  The first such method was used by Kuperberg in his algorithm
for the dihedral HSP\cite{Kuperberg:03a} (where it is listed as Proposition
9.1).  The basic idea of Kuperberg's method is as follows.  Suppose that we are
working on the tensor product of two Hilbert spaces, each of which is spanned
by a basis of the group elements $\{|g\rangle , g \in {\mathcal G}\}$.  Define
the following unitary operator on this tensor product space:
\begin{equation}
U_l(|a\rangle \otimes |b\rangle)=(|b^{-1}a\rangle \otimes |b\rangle)
\end{equation}
where $a,b \in {\mathcal G}$.  Then this operation takes left multiplication by
the direct product of the group to left multiplication on the right Hilbert
space.  In other words,
\begin{equation}
U_l [R_L(g) \otimes R_L(g)] U_l^\dagger = I \otimes R_L(g).
\end{equation}
Kuperberg uses this observation to perform {\em summand extraction}.  Summand
extraction works by taking two spaces which carry irreducible representations
of a group ${\mathcal G}$ and then performing quantum Fourier transforms (and
inverses) over ${\mathcal G}$ along with $U_l$ (we refer the reader to
\cite{Kuperberg:03a} for details.)  The net effect of these transforms is to
perform a partial Clebsch-Gordan transform.  In particular whereas in a full
Clebsch-Gordan transform, one has access to registers containing the total
irrep label $|\mu\rangle$, the space where the multiplicity acts $|w\rangle$,
and the space where the total irrep acts $|v\rangle$ (see Section
\ref{sec:multihscp}), in using this method the multiplicity register is not
directly accessible.  In other words there is no canonical basis which could be used to reveal the information stored in the multiplicity register.  Or, to put it differently, none of the algorithmic uses to which this method has been applied use the multiplicity register and therefore arbitrary basis for this information was defined.  The method of {\em summand extraction} is also explored by Moore, Russell and Sniady\cite{Moore:06a} where again the multiplicity register is not used.

Finally, a similar method for getting access to the irrep register and the
space where the irrep acts for a Clebsch-Gordan transform is described in an early  preprint version of \cite{Bacon:06d}, as well as in the the thesis of
Harrow\cite{Harrow:05a}.  This method uses a generalizes the quantum phase
estimation algorithm to non-Abelian groups.  For our purposes, just as in
summand extraction, these circuits can be used to enact a Clebsch-Gordan
transform in its entirety, but there has been no work describing how information in the multiplicity register can be accessed.

We can now put our Clebsch-Gordan transform into perspective.  While previous work has focused on the role of the irrep label register and the register where the representation lives, none has focused on the multiplicity space register.  Our circuit for a Clebsch-Gordan transform picks out a particular basis for this multiplicity space, and, as we shall show in the next section, this will lead to an efficient algorithm for the Heisenberg HSCP.

\section{The Clebsch-Gordan Transform over the Heisenberg Group and Solving
the
Heisenberg HSCP} \label{sec:algo}

We will now show that applying the Clebsch-Gordan transform over the Heisenberg
group allows us, with a little bit more work, to solve the Heisenberg HSCP
using two copies of the Heisenberg hidden subgroup state.

Consider the general procedure described in Section~\ref{sec:multihscp} for two
copies of the Heisenberg hidden subgroup state.  This procedure begins by
prepare two copies of the hidden subgroup states, perform a QFT over the
Heisenberg group on these states, measuring the irrep index, and throwing away
the registers where the left regular representation acts.  With probability
$(p-1)(p-2) \over p^2$ we will obtain a label corresponding to the $k_1$st
irrep, $\sigma_{k_1}$, in the first hidden subgroup and $k_2$st irrep,
$\sigma_{k_2}$, in the second hidden subgroup state with $k_1$ and $k_2$
satisfying $k_1+k_2 \neq 0~{\rm mod}~p$.  This probability is exponentially
close to $1$, so we can assume that we are exactly in this situation.  Further
we may assume that $k_1$ and $k_2$ are given with uniform probability assuming
that $k_1+k_2 \neq 0$.  In this case, the two registers where the right regular
representation acts contains the state
\begin{equation}
\rho_{k_1}({\mathcal A}_{i,j}) \otimes \rho_{k_2}({\mathcal A}_{i,j}).
\end{equation}

We will now apply the Clebsch-Gordan transform to this state.  Given that we
are dealing with two $p$ dimensional irreps, with $k_1+k_2 \neq 0{~\rm mod}~p$,
this is equivalent to applying the $W$ of Eq.~(\ref{eq:cgW}) to the state and
to adding the two irrep labels $k_1$ and $k_2$ together to obtain the new irrep
label, $k^\prime=k_1+k_2 \neq 0$.  Using Eq.~(\ref{eq:rhok}) we find that
\begin{eqnarray}
&&W (\rho_{k_1}({\mathcal A}_{i,j}) \otimes \rho_{k_2}({\mathcal A}_{i,j}))
W^\dagger \nonumber \\&&={1 \over p^2}\sum_{l_1,l_2,r_1,r_2 \in {\mathbb Z}_p}
 \omega^{k_1 (2^{-1}l_1(l_1-1)i+l_1 r_1i+ l_1 k_1 j)+k_2 (2^{-1}l_2(l_2-1)i+l_2
 r_2i+ l_2 k_2 j)}  \\ &&
|r_1+l_1-r_2-l_2\rangle  \langle r_1-r_2|  \otimes
|[k_1(r_1+l_1)+k_2(r_1+l_1)](k^\prime)^{-1}\rangle  \langle [k_1 r_1+k_2
r_2](k^\prime)^{-1}| \nonumber
\end{eqnarray}
Next, we measure the second of these registers since this register will contain
no information about which conjugate subgroup the state belongs to.  If we make
this measurement, then we will obtain outcome $|m\rangle$ with probability $p$
and the state remaining in the first register will be
\begin{equation}
{1 \over p}\sum_{l_1,u \in {\mathcal Z}_p} \omega^{k_1 2^{-1}l_1^2 (1+k_1
k_2^{-1}) i +k_1 l_1 i u}|u+l_1(1+k_2^{-1}k_1) \rangle \langle u|,
\end{equation}
where we have used the fact that only on diagonal elements of the second
register matter, so $k_1 l_1+k_2 l_2=0 ~{\rm mod}~p$, and relabeled
$u=r_1-r_2$.  Further relabeling this sum by $s_1=u+l_1(1+k_2^{-1} k_1)$ and
$s_2=u$ this becomes
\begin{equation}
{1 \over p} \sum_{s_1,s_2 \in {\mathbb Z}_p} \omega^{i 2^{-1} {k_1 k_2 \over
k_1+k_2} (s_1^2-s_2^2)} |s_1\rangle \langle s_2|.
\end{equation}

At this point we can begin to see how to solve the Heisenberg HSCP by making a
particular measurement on this state.  In particular the above density matrix
corresponds to the pure state
\begin{equation}
{1 \over \sqrt{p}} \sum_{s \in {\mathbb Z}_p} \omega^{i 2^{-1}{k_1 k_2 \over
k_1+k_2} s^2} |s\rangle. \label{eq:cgout}
\end{equation}
Suppose, now, that we had a transform $U_2$ which acted upon an equal
superposition of the two square roots (done modulo $p$) of a number in the
computational basis, ${1 \over \sqrt{2}}(|\sqrt{t}\rangle+|-\sqrt{t}\rangle)$
and produced the square number $|t\rangle$, and if it acts on the state
$|0\rangle$ it produces the state $|0\rangle$.  Applying this transform to the
our state results in
\begin{equation}
{\sqrt{2 \over p}}\sum_{v \in {\mathbb Z}_p|v~{\rm is~square}\neq 0} \omega^{i
2^{-1}{k_1 k_2 \over k_1+k_2} v} |v\rangle+{1 \over \sqrt{p}}|0\rangle
\end{equation}
It is convenient to rewrite this state as
\begin{equation}
{\sqrt{2 \over p}}\left[\sum_{v \in {\mathbb Z}_p} \omega^{ a v}
|v\rangle-\sum_{v \in {\mathbb Z}_p|v~{\rm not~square}} \omega^{ a v}
|v\rangle\right]+{1-\sqrt{2} \over \sqrt{p}}|0\rangle
\end{equation}
where $a=i{k_1 k_2  \over 2(k_1+k_2)}$.  Now performing an inverse QFT over
${\mathbb Z}_p$ results in the state
\begin{equation}
\sqrt{2}|a\rangle-{\sqrt{2} \over p} \sum_{v \in {\mathbb Z}_p|v~{\rm
not~square}}\sum_{x \in {\mathbb Z}_p} \omega^{av-vx}|x\rangle+{1 -\sqrt{2}
\over p}\sum_{x \in {\mathbb Z}_p}|x\rangle
\end{equation}
If we now measure this state, then the probability that we obtain the outcome
$a$ is
\begin{eqnarray}
Pr(a)=\left[ \sqrt{2}- \sum_{v \in {\mathbb Z}_p|v~{\rm not~square}}{ \sqrt{2}
\over p}+{1 - \sqrt{2} \over p} \right]^2 =\left[{1 \over \sqrt{2}}+(1-{1 \over
\sqrt{2}}) {1 \over p} \right]^2={1 \over 2}+O({1 \over p^2})
\end{eqnarray}
Thus we see that with probability approximately ${1 \over 2}$ a measurement
will result in the outcome $|a\rangle$.  Since we know $k_1$ and $k_2$, we can
easily invert $a=i {k_1 k_2 \over k_1+k_2}$ and find $i$, the label of the
hidden subgroup conjugacy.

Thus, assuming that we can efficiently implement $U_2$ we have shown how to
efficiently solve the HSCP for the Heisenberg.  How do we implement $U_2$
efficiently?  This can be done via an idea explained in the original Heisenberg
hidden subgroup algorithm of Bacon, Childs, and van Dam\cite{Bacon:05a}.  In
particular we note that there exists an efficient deterministic classical
algorithm which can compute the two square roots of a number modulo $p$.  In
particular this algorithm can work with a control bit labeling which of the two
square roots is returned: $V|x\rangle |b\rangle=|(-1)^b \sqrt{x}\rangle
|b\rangle$ where $\sqrt{x}$ is a canonical square root.  Then consider running
this algorithm backwards, with an input on the control bit being $|b\rangle={1
\over \sqrt{2}}(|0\rangle+|1\rangle)$.  After this operation, perform a
Hadamard on the control qubit and measure this register.  If the outcome is
$|0\rangle$, the applied transform on the non-control register will take ${1
\over \sqrt{2}}(|\sqrt{x}\rangle+|-\sqrt{x}\rangle)$ to $|x\rangle$ as desired.
This will occur with probability ${1 \over 2}$.  Thus we find that we can
efficiently implement $U_2$ with probability ${1 \over 2}$ such that our full
algorithm succeeds with probability ${1 \over 4}+O({1 \over p^2})$.

We have thus shown that the Clebsch-Gordan transform plus some quantum post
processing can be used to efficiently solve the Heisenberg HSCP and hence solve
the full Heisenberg HSP.

\subsection{Comparison to the PGM Heisenberg HSP Algorithm}

Above we have shown how to arrive at an efficient quantum algorithm for the
Heisenberg HSP by focusing on the Heisenberg HSCP and using the extra symmetry
of this state to motivate the use of a Clebsch-Gordan transform to solve the
problem.  How does this compare with the algorithm derived by Bacon, Childs,
and van Dam\cite{Bacon:05a}?  We will see below that we have, in effect,
derived a portion of this algorithm.

Here we will briefly recap the hidden subgroup algorithm of \cite{Bacon:05a}
for the Heisenberg group which was derived using the PGM formalism.  The first
step begins with the standard preparation of the hidden subgroup state.  One
then performs a quantum Fourier transform over ${\mathbb Z}_p^2$ on the
registers corresponding to the $y$ and $z$ values of the group element
$(x,y,z)$, and measuring these registers, obtaining with uniform probability
over ${\mathbb Z}_p$ a value for $y$ and $z$.  After this one is left with the
state,
\begin{equation}
|i,j,y,z\rangle={1 \over \sqrt{p}}\sum_{b \in {\mathbb Z}_p} \omega^{b j y
+2^{-1} b(b-1)i y +bj z} |b\rangle, \label{eq:pgmstate}
\end{equation}
for random uniform $y,z \in {\mathbb Z}_p$.  When we have two copies of this
state, we obtain four uniform random variables, $y_1,z_1,y_2,z_2 \in {\mathbb
Z}_p$ along with the state
\begin{equation}
|i,j,y_1,z_1\rangle \otimes |i,j,y_2,z_2\rangle={1 \over p} \sum_{b_1,b_2 \in
{\mathbb Z}_p} \omega^{j (b_1 y_1+b_2 y_2) +2^{-1}i( b_1(b_1-1) y_1+b_2
(b_2-1)y_2)  +j(b_1 z_1+b_2 z_2)}|b_1,b_2\rangle
\end{equation}
If we rewrite this state in terms of the variables
\begin{eqnarray}
u&=&b_1 y_1+b_2 y_2 \nonumber \\
v&=&2^{-1}b_1(b_1-1) y_1+2^{-1} b_2 (b_2-1)y_2+b_1 z_1+b_2 z_2
\label{eq:quad}
\end{eqnarray}
the two copy state can be expressed as
\begin{equation}
{1 \over p} \sum_{b_1,b_2 \in {\mathbb Z}_p} \omega^{u j +v i} |b_1,b_2\rangle
\end{equation}
One then notes that if one could enact a transformation which takes the an
equal superposition of the two solutions to the quadratic equations defined in
Eq.~(\ref{eq:quad}) to their respective values of $u$ and $v$, then we could
transform the above state into
\begin{equation}
{1 \over p} \sum_{u,v \in {\mathbb Z}_p|u,v ~{\rm
solve~the~quadratic~equation}} \omega^{u j +v i} |u,v\rangle
\end{equation}
from which a QFT over ${\mathbb Z}_p^2$ will immediately reveal $i$ and $j$
with probability ${1 \over 2}$ (since only half of the values of $u$ and $v$
will arise in solving the quadratic equations.)  Exactly such a transform which
can take the superposition over two solutions to the $u$ and $v$ values which
can be enacted using the same basic idea as that used to construct $U_2$
efficiently above, but instead of taking the two square roots to the square, a
more complicated quadratic equation must be solved.  We refer the reader to
\cite{Bacon:05a} for details of this construction.

From the above description of the PGM derived quantum algorithm for the
Heisenberg HSP we can use the insights we have gained in solving this using a
Clebsch-Gordan transform to explain the structure of the above algorithm.
First note that by performing a QFT over ${\mathbb Z}_p^2$, measuring the
outcome and producing the state in Eq.~(\ref{eq:pgmstate}) we are, in effect,
doing something very near to that of the QFT over the Heisenberg group.  Notice
however that in the original PGM state, we keep around two indices $y$ and $z$.
However it is easy to see that if we average the pure states in
Eq.~(\ref{eq:pgmstate}) over $y$, we obtain another pure state
\begin{equation}
|i,j,y\rangle={1 \over \sqrt{p}}\sum_{b \in {\mathbb Z}_p} \omega^{2^{-1}
b(b-1)i y +bj y} |b\rangle.
\end{equation}
In other words the $y$ register contains no information about the hidden
subgroup.  This is like the register containing the left regular representation
not containing information about the hidden subgroup.  Thus in the two copy
state we could have equally well have dealt with the state
\begin{equation}
{1 \over p} \sum_{b_1,b_2 \in {\mathbb Z}_p} \omega^{u j +w i} |b_1,b_2\rangle
\end{equation}
where
\begin{eqnarray}
u&=&b_1 y_1+b_2 y_2 \nonumber \\
w&=&2^{-1}b_1(b_1-1) y_1+2^{-1} b_2 (b_2-1)y_2.
\end{eqnarray}
We see that $y_1$ and $y_2$ serve as the $k_1$ and $k_2$ irreps labels (with an
exception occurring when $k_1=k_2=0$.)

Next we note that, following our observation about how to perform the
Clebsch-Gordan transform on the Heisenberg group, we can perform a basis
change
\begin{equation}
\sum_{s,t \in {\mathbb Z}_p} |s-t\rangle  \langle s| \otimes |(s y_1 +t
y_2)(y_1+y_2)^{-1} \rangle \langle t|,
\end{equation}
assuming $y_1+y_2 \neq 0$.  This transforms $|i,j,y_1\rangle |i,j,y_2\rangle$
into
\begin{equation}
{1 \over p} \sum_{b_1,b_2 \in {\mathbb Z}_p} \omega^{u j +w i} |b_1-b_2,(y_1
b_1+y_2 b_2)(y_1+y_2)^{-1}\rangle.
\end{equation}
In the Clebsch-Gordan method we can measure the register of the irrep label.
In the PGM case this corresponds to performing the above transform and then
measuring the second register.  Let this second register measurement outcome
yield value $m$.  This will produce the state
\begin{equation}
{1 \over \sqrt{p}} \sum_{r \in {\mathbb Z}_p} \omega^{wi}|r\rangle \otimes
|m\rangle,
\end{equation}
where $r=b_1-b_2$ and we can express $w$ as
\begin{equation}
w=2^{-1}\left[ {y_1 y_2 \over {y_1+y_2}} r^2+m^2(y_1+y_2)+m\right],
\end{equation}
which, factoring out the global phase dependent on $m$ is
\begin{equation}
{1 \over \sqrt{p}} \sum_{r \in {\mathbb Z}_p} \omega^{{y_1 y_2 \over 2
(y_1+y_2)}i r^2}|r\rangle \otimes |m\rangle.
\end{equation}
Which we see is exactly the state we obtained after the Clebsch-Gordan
transform, Eq.~(\ref{eq:cgout}).  Thus we see that the effect of the
Clebsch-Gordan transform is to transform the quadratic equation in
Eq.~(\ref{eq:quad}) into a form in which the resulting quadratic equation
contains only a quadratic term.  So, in quite a real sense, our derivation of
solving the HSCP using the Clebsch-Gordan transform leads directly to at least
part of the measurement for the HSP used in the PGM approach.  Thus it is best
to view our Clebsch-Gordan derivation as giving a representation theoretic
derivation of the PGM algorithm.  Note however that it does not give quite the
same derivation, since in the PGM based algorithm one obtains the full subgroup
while in the Clebsch-Gordan based algorithm we only obtain which set of
conjugate subgroups the hidden subgroup belongs to, using the hidden subgroup
algorithm for normal subgroups to completely identify the hidden subgroup.

\section{Conclusion}

By focusing on the HSCP instead of the HSP, we have been able to apply symmetry
arguments to multicopy algorithms which do not hold for the HSP.   These
symmetry arguments were exploited by performing a Clebsch-Gordan transform over
the Heisenberg group and then applying an appropriate post processing
measurement on the multiplicity register output of this transform.  The
algorithm we describe bears a great deal in common with the algorithm of Bacon,
Childs, and van Dam\cite{Bacon:05a}.  However, we believe that an important
insight into {\em why} the algorithm of Bacon, Childs, and van Dam works has
been discovered.  In particular, instead of relying on the optimal measurement
criteria, we instead see that focusing on the symmetry of the multicopy HSCP
leads to a change of basis which great facilitates solving the Heisenberg HSCP.
In a real sense this implies that the Clebsch-Gordan transform over the
Heisenberg group naturally arises for our problem.  We believe that this is a
new insight whose significance bears further investigation for other
non-Abelian HSPs.

At this point it is useful to draw an analogy with a previous stage in quantum algorithms research.  For a long time it was known that a quantum Fourier transform could be used on hidden subgroup states without destroying any of their coherence.  However quantum Fourier transforms do not have a canonical basis choice for their registers and thus the question investigated by researchers was whether a good choice of basis exists for solving the hidden subgroup problem.  The Clebsch-Gordan transform similarly does not have a canonical basis choice for the multiplicity register.  We have shown, however, that for the case of the Heisenberg group, a particular choice of basis, {\em can} be used to eventually solve the hidden subgroup problem.  The work performed on choosing a proper basis for the quantum Fourier transform led to the result of Hallgren {\em et al}\cite{Hallgren:06a} that multi-register measurements are needed to solve the HSP.  We are hopeful that understanding the proper basis choice for the multiplicity space register of the Clebsch-Gordan transform will not yield such negative results and this work represent evidence in favor of this view.

Numerous obvious open problems arise from the above investigations.  An immediate question is for what other groups can Clebsch-Gordan transforms be used to efficiently solve the HSCP or HSP?  Another important open issue is over what groups can the Clebsch-Gordan transform be efficiently enacted in such a way as there is some natural basis choice for the multiplicity space register.  In this
regard, the subgroup adapted basis techniques of Bacon, Harrow, and Chuang\cite{Bacon:06d,Bacon:06e}
along with those of Moore, Rockmore, and Russell\cite{Moore:04a}, should be of
great help.  Finally, an interesting question is whether the HSP and HSCP are
(classical or quantum) polynomial time equivalent to each other.  While this is known to be true for certain finite groups, the general case remains open.  Of
particular significance is this question for the dihedral groups when the group order is not smooth.

\nonumsection{Acknowledgements}

DB is supported under ARO/NSA quantum algorithms grant number
W911NSF-06-1-0379, NSF grant number 0523359, and NSF grant number 0621621.  DB
would like to acknowledge Andrew Childs for suggesting many improvements in an
earlier draft of this paper and Thomas Decker for discussions on the Heisenberg
group.  DB would also like to thank John Preskill for repeatedly asking DB
whether Clebsch-Gordan transforms are actually useful for quantum algorithms.

\nonumsection{References}

\end{document}